\documentclass[prl,aps,10pt,twocolumn,superscriptaddress,showpacs]{revtex4-1}

\usepackage{epsfig}
\usepackage{amsmath}
\usepackage{graphicx}
\usepackage{graphics}
\usepackage{float}
\usepackage{amssymb}
\usepackage[usenames,dvipsnames]{color}
\usepackage{natbib}
\usepackage{epstopdf}
\usepackage{verbatim}
\usepackage{wrapfig}
\usepackage{subfigure}

\newcommand{\ket}[1]{\left| #1\right\rangle}
\newcommand{\ip}[2]{\langle#1|#2\rangle}

\newcommand{\jn}[1]{~{\bf #1}}
\newcommand{\opav}[3]{\left\langle #1 | #2 | #3 \right\rangle}

\newcommand{\beq}{\begin{equation}}
\newcommand{\eeq}{\end{equation}}

\begin{document}

\title{Zeno and Anti-Zeno Effects on Dephasing}

\author{Adam Zaman Chaudhry}
\email{adamzaman@gmail.com}
\affiliation{Department of Electrical and Computer Engineering, National University of Singapore, 4 Engineering Drive 3, Singapore 117583, Singapore}
\author{Jiangbin Gong}
\email{phygj@nus.edu.sg}
\affiliation{Department of Physics and Center for Computational
Science and Engineering, National University of Singapore, Singapore 117542,
Singapore}
\affiliation{NUS Graduate School for Integrative Sciences
and Engineering, Singapore 117597, Singapore}

\begin{abstract}
Quantum Zeno and anti-Zeno effects on pure dephasing are studied using exactly solvable microscopic models. The crossover between these two opposite effects is investigated. The case of a single two-level system undergoing dephasing is already different from the previously studied population decay problem, even without taking into account any back action from the environment. For many two-level systems interacting with a common environment, multiple transitions between Zeno and anti-Zeno regimes are predicted. Finally, if the system-environment coupling strength is not weak, we show that the nontrivial evolution of the environment between measurements can considerably alter the quantum Zeno and anti-Zeno effects.

\end{abstract}

\date{\today}

\maketitle

The quantum Zeno effect (QZE) slows down the evolution of a quantum system under rapidly repeated measurements \cite{Sudarshan1977}.   QZE has been proposed to freeze, or at least suitably confine, the evolution of a quantum state and continues to be a topic of great theoretical and experimental interest \cite{QZDreferences}.
 However, when the measurements are not frequent enough, they may actually accelerate quantum transitions, an effect dubbed as the
the quantum anti-Zeno effect (QAZE) \cite{KurizkiNature2000}.
Both QZE and QAZE have been investigated in different contexts, such as localized atomic systems \cite{RaizenPRL2001}, superconducting current-biased Josephson junctions \cite{BaronePRL2004}, disordered spin systems \cite{YamamotoPRA2010} and nanomechanical oscillators \cite{BennettPRB2010}.  To date studies of QZE-QAZE transitions have focused on the population decay of a quantum system interacting with an environment where the measurement in action is to determine the population of an excited state \cite{KurizkiNature2000, RaizenPRL2001, BaronePRL2004, YamamotoPRA2010, BennettPRB2010, KoshinoPhysRep2005, ManiscalcoPRL2006, SegalPRA2007, ZhengPRL2008, AiPRA2010, ThilagamJMP2010, ThilagamJCP2013}. In this scenario, the interplay between QZE and QAZE has been linked with an overlap integral between the spectral density of the environment and a measurement-induced level width \cite{KurizkiNature2000}.
However, experimental demonstrations of the QZE-QAZE crossover \cite{RaizenPRL2001} are in general demanding due to the large measurement rates required \cite{NoteonZenotime}, thus motivating the development of frontier technologies.

Here we are concerned with QZE and QAZE on ``dephasing" \cite{BPbook, Weissbook}.   Fighting against dephasing is a crucial and challenging step towards practical implementations of emerging quantum technologies.  Because dephasing can occur much faster than population decay, the crossover between QZE and QAZE may emerge on even shorter time scales and remains to be carefully examined. Indeed, using a microscopic exactly solvable pure-dephasing model, we show that QZE and QAZE therein are different from their parallel population-decay problems.  Extending our considerations to a collective dephasing model, we predict multiple QZE-QAZE transitions. Finally, measurements not only disturb the system, but also project the environment onto non-equilibrium states \cite{KurizkiNJP2010}.  The nontrivial evolution of the environment between measurements due to system-environment correlations is shown to be important for QZE and QAZE when the system-environment coupling is not weak.  In this sense,  QAZE becomes an outcome of the disturbance of repeated measurements to both the system of interest and its environment.
These results are hoped to stimulate future experiments.

{\it Pure-dephasing of a single two-level system} -- We start with a pure-dephasing spin-boson Hamiltonian \cite{BPbook}
$H = H_S + H_B + H_{SB}$,
where $H_S = \frac{\omega_0}{2}\sigma_z$, $H_B = \sum_k \omega_k b_k^\dagger b_k$, and $H_{SB} = \sigma_z \sum_k (g_k^* b_k + g_k b_k^\dagger)$.  Throughout, we work in dimensionless units and set $\hbar = 1$. In the $\sigma_z$ eigenbasis defined by $\sigma_z\ket{e} = \ket{e}$ and $\sigma_z\ket{g} = -\ket{g}$, the diagonal elements of the system's reduced density matrix do not change with time, and the off-diagonal element, assuming throughout that the initial system-environment state is a product state with the environment in a thermal state, is given by $
[\rho(t)]_{eg} = [\rho(0)]_{eg} e^{-i\omega_0 t} e^{-\gamma(t)}$,
where
\begin{equation}
\gamma(t) = 4\sum_k \frac{|g_k|^2}{\omega_k^2} [1 - \cos(\omega_k t)] \coth\left( \frac{\beta \omega_k}{2}\right)
\end{equation}
depicts the environment-induced dephasing, i.e., the loss of coherence between states $\ket{e}$ and $\ket{g}$.
We are mainly interested in QZE and QAZE within the dephasing time scale, i.e., $e^{-\gamma(t)}\approx 1$.

We consider an initial equal-weight superposition state $\ket{\psi} = \frac{1}{\sqrt{2}}(\ket{e} + \ket{g})$ at time $t=0$ (extension to an arbitrary superposition state is trivial).
For now let us neglect any measurement-induced disturbance to the environment (at least valid for weak system-environment coupling).  $N$ repeated measurements of the projector $P_{\psi}=|\psi\rangle\langle\psi|$, with equal time interval $\tau$, are now applied, but before each measurement, we apply the rotation $U_R(\tau) = e^{iH_S \tau}$ which removes the system evolution induced by $H_S$ itself.  Because the survival probability $S$ would be just the $N\textsuperscript{th}$ power of the survival probability associated with one measurement, it is convenient to write $S\equiv e^{-\Gamma(\tau) t_N}$, with $t_N=N\tau$ and $1/\Gamma(\tau)$ being an {\it effective} lifetime of the initial superposition state that depends on the measurement interval $\tau$. One then obtains
\begin{equation}
\Gamma(\tau) = -\frac{1}{\tau} \ln \lbrace 1 -\frac{1}{2}[1 - e^{-\gamma(\tau)}]\rbrace.
\label{gammaeq}
\end{equation}
Note that $\Gamma$ obtained above is independent of $N$, which is a manifestation of our assumption that the measurements do not disturb the environment.
QZE on dephasing becomes obvious if sufficiently small measurement interval $\tau$ is considered.
In such cases $\gamma(\tau) \approx 2y\tau^2$, with $y = \sum_k |g_k|^2 \coth \left(\beta \omega_k/2 \right)$, and we obtain $
\Gamma(\tau) \approx a\tau$,
with $a=y$.
That is, as expected, a very
small $\tau$ leads to a vanishing $\Gamma$ and hence a frozen initial state.

\begin{figure}[b]
\centering
\includegraphics[scale = 0.5]{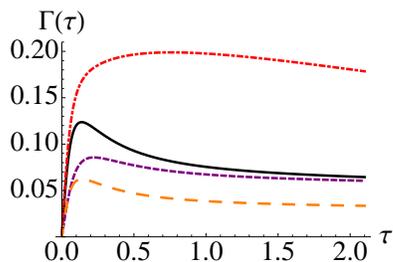}
\caption{(color online) Behavior of $\Gamma(\tau)$ as measurement interval $\tau$ is varied. The environment is assumed to have an Ohmic spectral density.
 We have set $G = 0.01$, $\omega_c = 15$, and $\beta = 1$ for the solid (black) curve, for which we obtain $\gamma(\tau = 5) = 0.65$. For the other curves, the parameters are the same, except that for the dashed (purple) curve, $\omega_c = 10$, while for the dot-dashed (red) curve, $\beta = 0.25$, and for the long-dashed (orange) curve, $G = 0.005$. For the values of $\beta$ and $\omega_c$ considered here, if $\omega_c$ is in the GHz regime, then the temperature is in the mK regime, the operating temperature for superconducting qubits \cite{NoriRepProg2011}.}
\label{Gammaexamplefig}
\end{figure}

Our pure-dephasing model affords a general expression of $\Gamma(\tau)$ in Eq.~(\ref{gammaeq}). One might wonder if QAZE can be captured by  Eq.~(\ref{gammaeq}) at all.  To that end we show typical features of $\Gamma(\tau)$ in Fig.~\ref{Gammaexamplefig}, where we model the environment using an Ohmic spectral density. That is,
$\sum_k|g_k|^2 \rightarrow \int J(\omega) d\omega$, with $J(\omega) = G\omega e^{-\omega/\omega_c}$ and $\omega_c$ is the cutoff frequency of the environment.
  One immediate observation is that $\Gamma(\tau)$ for all the shown examples has a peak structure.  As $\tau$ increases beyond a certain value,  denoted by $\tau_{\text{Z-AZ}}$,  $\Gamma(\tau)$ reverses its qualitative dependence on $\tau$ (i.e., smaller measurement interval now means larger $\Gamma$ and hence less survival probability).  In this sense, such a peak structure clearly indicates a (local) QZE-QAZE transition \cite{Noteonlocal}. Comparing between the shown examples, it is seen that properties of the environment, such as its cutoff frequency, its temperature, and the system-environment coupling, can all influence the value of $\tau_{\text{Z-AZ}}$. For example, an increase in the environment temperature can significantly widen the QZE regime (comparing the top two curves in Fig.~\ref{Gammaexamplefig}).  Note also that for the shown examples,  $\tau_{\text{Z-AZ}}$ as the time scale for QZE-QAZE transitions is much shorter than the characteristic decoherence time scale.  This presents a potential challenge for experimental observations. However, in experimental demonstrations one may start from
a well-isolated quantum system and then add engineered weak dephasing into the system, thus synthesizing a system with a long dephasing time scale and hence a relatively larger $\tau_{\text{Z-AZ}}$.

To better digest the QZE-QAZE transitions one may perform higher-order expansions with respect to $\tau$, e.g., $\gamma(\tau) \approx 2y\tau^2 - \frac{z\tau^4}{6}$, with $z = \sum_k |g_k|^2 \omega_k^2 \coth \left( \frac{\beta \omega_k}{2} \right)$. We can then eventually write
\begin{equation}
\label{Gammatauexpansion}
\Gamma(\tau) \approx a\tau + b\tau^3,
\end{equation}
with $b = - y^2/2 - z/12 $. This approximation predicts that $\Gamma(\tau)$ exhibits a peak, a prediction borne out by the curves in Fig.~\ref{Gammaexamplefig}.   It is also interesting to compare $\Gamma(\tau)$ in Eq.~(\ref{Gammatauexpansion})  with a parallel expression in the standard population decay problem studied earlier \cite{KurizkiNature2000}.  As detailed in Supplementary Material \cite{Supple}, for the spontaneous emission of an excited state (treated under the rotating-wave approximation) with measurement interval $\tau$, the modified decay constant for a zero temperature environment is given by $\tilde{\Gamma}(\tau) = \tau \sum_k |g_k|^2 \text{sinc}^2 [(\omega_k - \omega_0)\tau/2]$, where $\omega_0$ and $\omega_k$ carry the same meaning as in this work.   Also expanding $\tilde{\Gamma}(\tau)$ to the third order of $\tau$, we have $\tilde{\Gamma}(\tau) = \tilde{a}\tau + \tilde{b}\tau^3$, with $\tilde{a}= \sum_k |g_k|^2$ and $\tilde{b}=-\sum_k|g_k|^2 (\omega_{k} - \omega_0)^2/12$. Remarkably, although our model can be transformed to a population transition problem with two degenerate levels \cite{Supple} (with a counter-rotating term), ${\Gamma}(\tau)$ obtained in Eq.~(\ref{Gammatauexpansion}) differs from $\tilde{\Gamma}(\tau)$ even after setting $\beta=\infty$ and $\omega_0=0$: $a=\tilde{a}$ but $b=\tilde{b}-\left(\sum_k|g_k|^2\right)^2/2$.  Thus, even in cases with small $\tau_{\text{Z-AZ}}$ and even without
considering the back action of the environment, QZE and QAZE for pure dephasing are already different from the previously studied population decay problem \cite{Supple}.

\begin{figure}[t]
\centering
\mbox{\subfigure{\includegraphics[scale = 0.4]{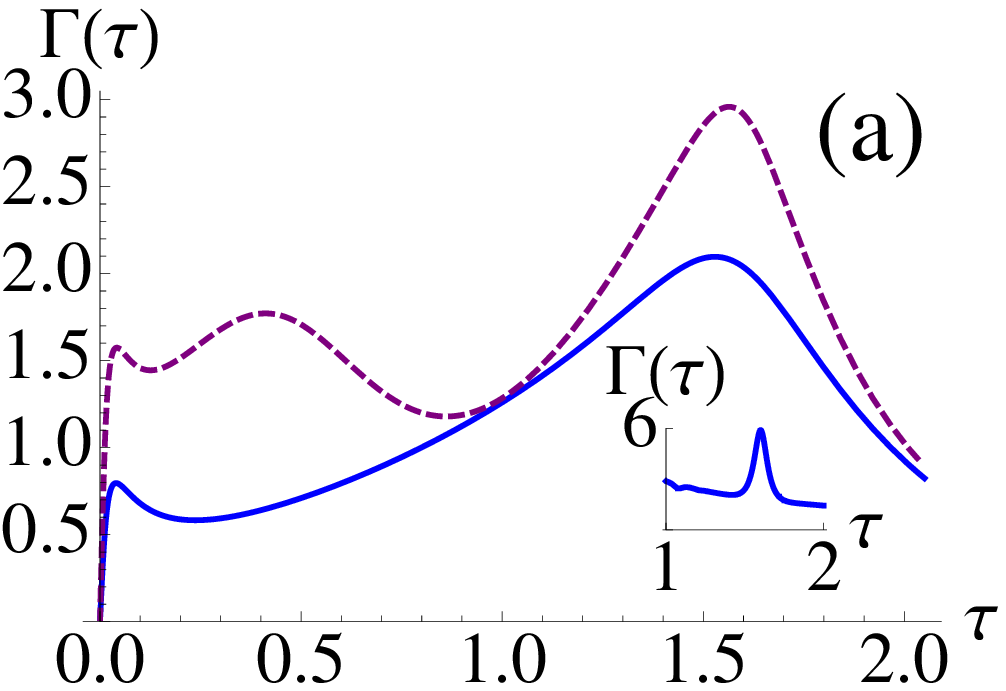}\label{bigJGamma} }\quad
\subfigure{\includegraphics[scale = 0.4]{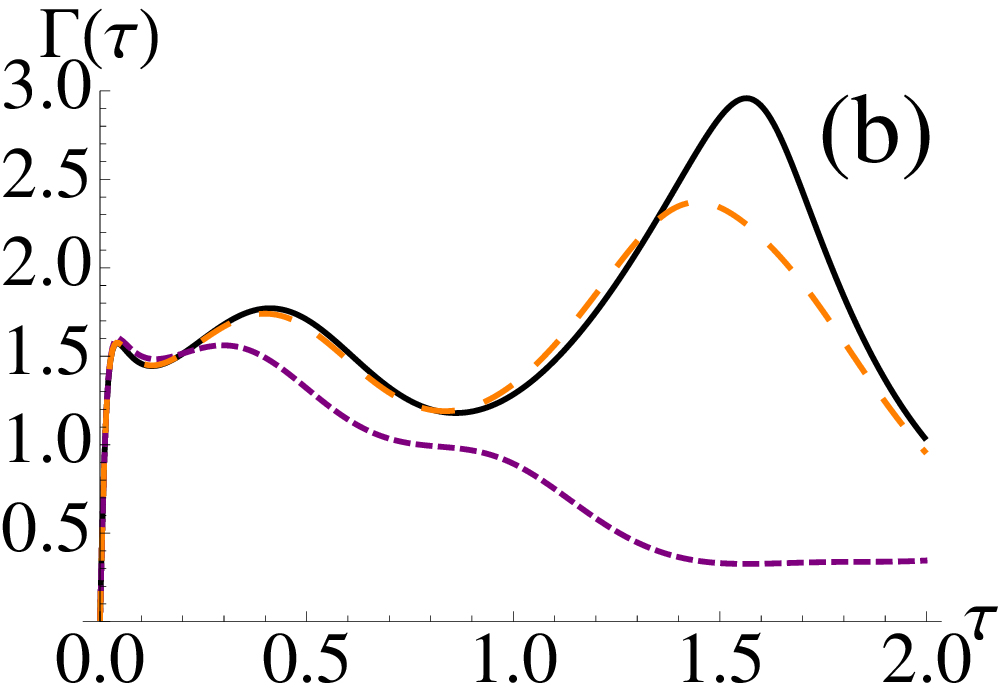}\label{dephasinganddissipationJ2} }}
\caption{(color online) (a) $\Gamma(\tau)$ vs $\tau$  for $J = 1$ (solid, blue) and $J = 2$ (dashed, purple) and $J = 50$ (inset, which shows that an additional peak still exists at about $\tau=1.6$). We have used $G = 0.01$, $\omega_c = 50$, and $\beta = 1$, and throughout we take $\theta = \pi/2$ and $\phi = 0$.
 (b) $\Gamma(\tau)$ for $J = 2$  with $G = 0.01$, $\omega_c = 50$, $\omega_0 = 0.1$ and $\delta = 0$ (solid, black), $\delta = 0.1$ (long-dashed, orange) and $\delta = 1$ (dashed, purple).}
\end{figure}

{\it Multiple QZE-QAZE transitions in a collective dephasing model} -- Our pure dephasing setup allows for a direct extension to a collective dephasing problem, in which many two-level systems interact with a common environment.   The system Hamiltonian then becomes $H_S = \omega_0 J_z$, and the system-envrionment coupling Hamiltonian becomes
$H_{SB}=2J_z \sum_k (g_k^* b_k + g_k b_k^\dagger)$ \cite{VorrathPRL2005},
where $J_z$, a collective spin operator, is half of the sum of all $\sigma_z$ operators for the spins.
 This model is also relevant to two-component Bose-Einstein condensates \cite{GrossNature2010, RiedelNature2010, KurizkiPRL2011}. In the eigenbasis of $J_z$, the system density matrix elements
are found to be
\begin{equation*}
[\rho(t)]_{mn} = [\rho(0)]_{mn} e^{-i\omega_0 (m - n)t} e^{-i\Delta(t) (m^2 - n^2)} e^{-\gamma(t) (m - n)^2}.
\end{equation*}
Here $\Delta(\tau) = 4\sum_k \frac{|g_k|^2}{\omega_k^2} [\sin(\omega_k \tau) - \omega_k \tau]$ describes the indirect interaction between the two-level systems due to their interaction with a common environment.  Such environment-induced indirect interaction leads to `phase diffusion' \cite{CastinPRA1997},  which degrades the reduced single-particle coherence. For vanishingly small time $t$, however, $\Delta(t) \approx 0$. On the other hand, as $t$ increases,  the effect of $\Delta(t)$ becomes more pronounced: it leads to revivals in the survival probability in the absence of any measurement \cite{Supple}.  This makes it interesting to investigate
 what happens if repeated measurements are turned on.  In particular,
we take the initial state of the system as a standard SU(2) coherent state \cite{ArecchiPRA1972}
\begin{equation*}
\ket{\varsigma,J} = (1 + |\varsigma|^2)^{-J} \sum_{m = -J}^J \sqrt{\binom{2J}{J + m}} \varsigma^{J + m} \ket{J,m},
\end{equation*}
where $\varsigma = e^{i\phi} \tan(\theta/2)$, $\phi$ and $\theta$ parametrizing the state on the Bloch sphere, states $|J,m\rangle$ are in the angular momentum notation, with $J_z|J,m\rangle = m|J,m\rangle$ and $J$ being half of the number of two-level systems. For $J=1/2$ we return to the single two-level case.
Again assuming that the state of the environment is not affected by measurements,
the inverse lifetime $\Gamma(\tau)$ is found to be
\begin{align}
\Gamma(\tau) = &- \frac{1}{\tau} \ln\Bigg\lbrace\left[ \frac{|\varsigma|}{(1 + |\varsigma|^2)}\right]^{4J} \sum_{m,n} |\varsigma|^{2(m + n)} \binom{2J}{J + m} \times \notag \\
&\binom{2J}{J + n}  e^{-i \Delta(\tau) (m^2 - n^2)} e^{-\gamma(\tau)(m - n)^2}\Bigg\rbrace.
\end{align}
Some computational examples are shown in Fig.~\ref{bigJGamma}.  It is seen that $\Gamma(\tau)$ now in general has multiple peaks or a recurrence of Zeno and anti-Zeno regimes at intermediate measurement intervals. Qualitatively, this may be understood as a coincidental matching between the measurement timings and the oscillatory nontrivial dynamics arising from the indirect interaction mediated by the environment \cite{Supple}. The existence of multiple peaks can be significant for experiments, because now a (local) QZE-QAZE transition may be also observed using a relatively large measurement interval $\tau$.

 To check the robustness of the multiple QZE-QAZE transitions to dissipation, we change the system Hamiltonian from $H_S = \omega_0 J_z$ to $H_S = \omega_0 J_z + \delta J_x$.  This problem can no longer be exactly solved so we use the non-Markovian master equation \cite{KurizkiPRL2004}
\begin{align}
\frac{d\rho(t)}{dt} &= i[\rho(t), H_S] \, + \notag \\
&\int_0^t dt' \, \Bigl\lbrace [\bar{F}(t')\rho(t), F] C(t') + \text{h.c.}] \Bigr\rbrace,
\end{align}
where we have written $H_{SB} = F \otimes B$ with $F = 2J_z$, $\bar{F}(t') = U_S(t')FU_S^\dagger (t')$ where $U_S(t')$ is the unitary time-evolution operator corresponding to the Hamiltonian $H_S$, $B = \sum_k (g_k b_k + g_k^* b_k^\dagger)$, and $C(t')$ is the environment correlation function.  This master equation yields the survival probability at time $\tau$ and $\Gamma(\tau)$ can then be found. As shown in Fig.~\ref{dephasinganddissipationJ2}, even when the system is far from a pure-dephasing case (for $\delta$ comparable with $\omega_0$), the multiple QZE-QAZE transitions survive.

{\it System-environment correlation effects on QZE-QAZE transitions} ---So far our discussions are based on the assumption that the state of the environment is not disturbed by the measurements: it is assumed to always be in the thermal equilibrium state after each measurement on the system.  This assumption, though made very often, becomes invalid if the system-environment coupling is not weak \cite{KurizkiNJP2010,adamCJC}.  Physical contexts with relatively strong system-environment coupling include superconducting qubits \cite{superconducting} and atom-cavity systems \cite{atom-cavity}.  Indeed, the system and its environment can get considerably correlated, and such system-environment correlations have recently been widely studied \cite{GessnerNatPhys2014,taka1}. As a result of the correlations,  when each measurement projects the system back to the initial density matrix $P_{\psi}=|\psi\rangle\langle\psi|$, it also re-prepares a new state for the environment.  For example, let $\rho_{\text{tot}}(0)$ be the initial state of the system plus the environment.  Then, after the the first measurement (along with a unitary rotation $U_R(\tau)$ to remove system's own evolution), the state of the environment is given by $\rho_e(\tau)=\langle \psi| U(\tau) \rho_{\text{tot}}(0) U^{\dagger}(\tau)|\psi\rangle/Z_1$, where $Z_i$ represents the normalization factor after $i$ measurements and
$U(\tau) = U_R(\tau) U_{\text{tot}}(\tau)$, where $U_{\text{tot}}(\tau)$ is the unitary evolution operator for the system and the environment as a whole.
After the second measurement, the environment state becomes proportional to $\langle \psi| U(\tau) P_{\psi}\otimes \rho_e(\tau) U^{\dagger}(\tau)|\psi\rangle$.  Thus, in general the state of
the environment keeps changing throughout the process.  Just before the $N\textsuperscript{th}$ measurement,
the state of the whole is given by
\begin{align*}
&\rho_{\text{tot}}(t \rightarrow N\tau) = \notag \\ &\lbrace U(\tau)[P_{\psi}U(\tau)]^{N-1} \rho_{\text{tot}}(0) [U^\dagger(\tau)P_{\psi}]^{N-1}U^\dagger(\tau)  \rbrace /Z_{N-1},
\end{align*}
which can be used to compute the system's survival probability $S(t=t_N)$ with $t_N=N\tau$. Interestingly, a general recipe for finding the exact survival probability for arbitrary $J$ and arbitrary $N$ can be found \cite{Supple}.
As before, we may define the inverse lifetime though $S(t=t_N)=e^{-\Gamma t_N}$ for a given total time $t_N$.  But now the entire evolution history of the environment matters, which rules out the possibility of having $S$ as a simple $N\textsuperscript{th}$ power of some expression.
Thus $\Gamma$ now depends on $\tau$ as well as the number of repeated measurements $N$.   This fact also hints the emergence of previously unknown phenomena regarding QZE and QAZE.

\begin{figure}[t]
\centering
\mbox{\subfigure{\includegraphics[scale = 0.4]{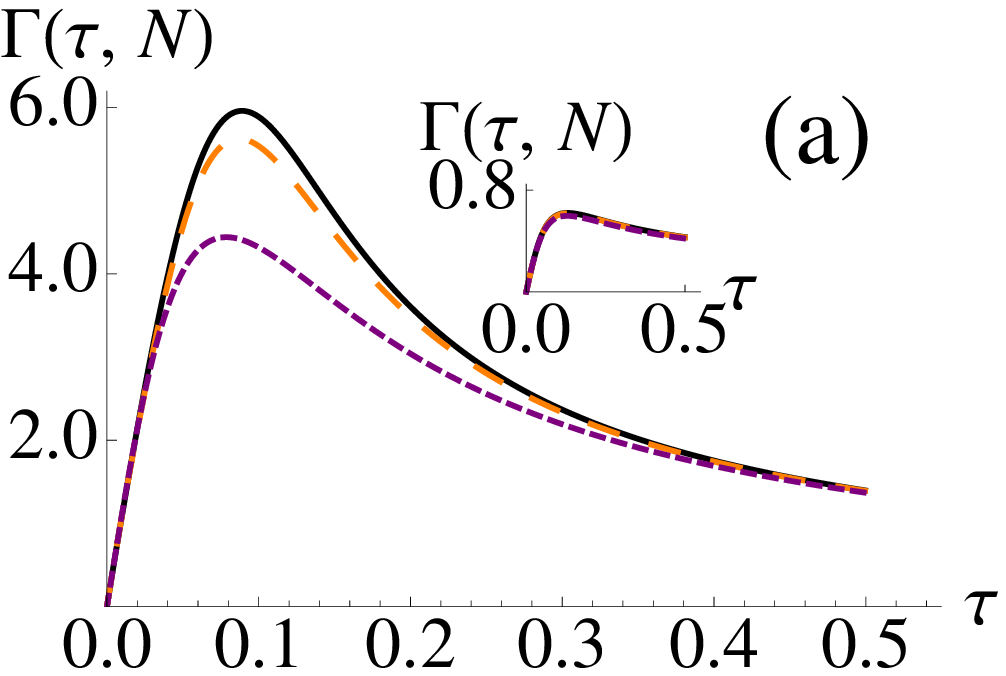}\label{Gammawithcorrelations}}\quad
\subfigure{\includegraphics[scale = 0.4]{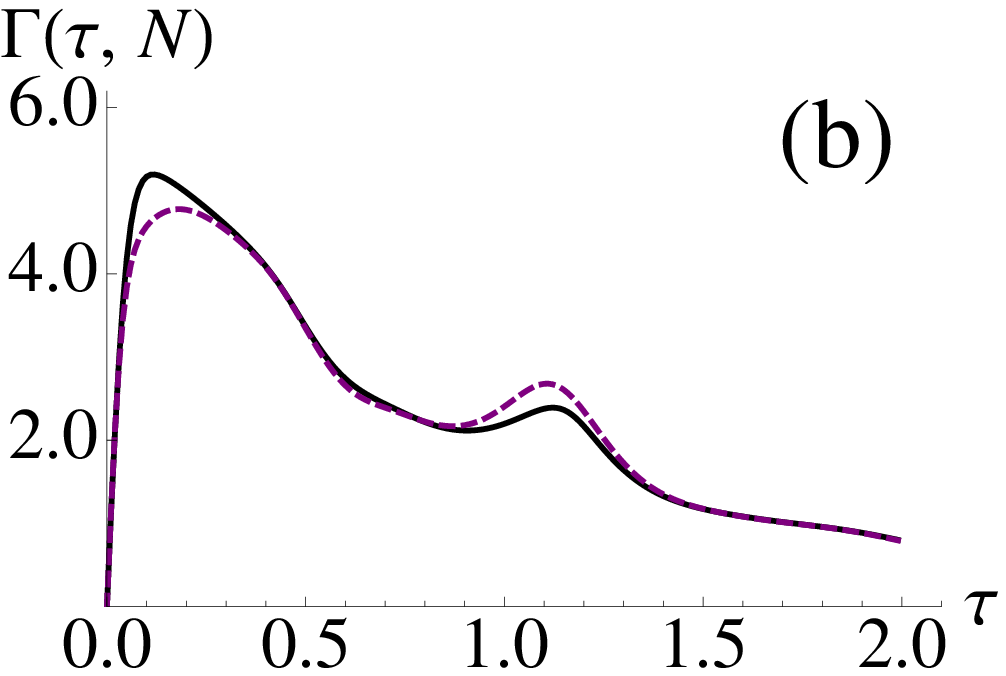}\label{GammalargeJwithcorrelations.eps} }}
\caption{(color online) (a) Effect of initial correlations on $\Gamma(\tau, N)$, for $J = 1/2$. For moderate coupling $G = 0.5$ (main figure), we have plotted $\Gamma(\tau, N)$ without system-environment correlations (dashed, purple), with correlations and $N = 3$ (long-dashed, orange) and $N = 5$ (solid, black). $\omega_c = 15$. $\tau_{\text{Z-AZ}}$ shifts from approximately $0.08$ to $0.09$ due to the correlations.  (Inset) Same as main figure, except that we now have $G = 0.05$. The lines now essentially overlap. (b) $\Gamma(\tau,N)$ without accounting for correlations (dashed, purple), and with correlations and $N = 3$ (solid, black), for $J = 5$ and $G = 0.05$. The first $\tau_{\text{Z-AZ}}$ changes from $0.18$ to $0.12$ due to the correlations, while the second $\tau_{\text{Z-AZ}}$ barely changes.}
\end{figure}

Our main findings are illustrated in Fig.~\ref{Gammawithcorrelations}, again using an Ohmic spectral density, with the conventional parameter $G$ representing the strength of system-environment coupling.
First, as a consistency check, the inset of Fig.~\ref{Gammawithcorrelations} shows that for a small $G=0.05$ and $J=1/2$, there is virtually no difference when accounting for the system-environment correlation effects.  By contrast, for cases of $G=0.5$ presented in the main panel,  the effect of system-environment correlations becomes appreciable for $N > 1$. Once the correlation effect is accounted for, the peak value of $\Gamma$ is increased and the location of its peak is shifted to a larger $\tau$. As the number of measurements is increased, this trend becomes even more pronounced.
Note also that, for $\tau$ at the respective crossover values, the survival probability after five measurements without correlations is approximately $0.17$, whereas it is only around $0.07$ with correlations, a big contrast of experimental relevance. This difference in the survival probability caused by the correlations reaches one order of magnitude for $N=10$ (with other parameters unchanged). The correlation effect can be further increased if we further increase $G$ or $N$.
Equally interesting, even if $G$ is small,  as the number of two-level systems increases ($J$ increases), the system-environment correlation effects start to influence the multiple QZE-QAZE transitions as well [see Fig.~\ref{GammalargeJwithcorrelations.eps}]. This is consistent with our earlier observation
 ~\cite{adamCJC} that the effect of one-time state preparation on the dynamics of an open quantum can be nontrivial for a large $J$ despite a small coupling parameter $G$.

{\it Physical realization} --- For a single two-level system, one can use any experimental realization of the spin-boson model \cite{Weissbook}. For instance, we can consider using a superconducting qubit \cite{superconducting}, which has the useful feature that both the energy bias and tunneling can be modified appropriately. In fact, the spin-boson model with tunable Ohmic dissipation can be realized for such qubits \cite{HurPRB2012}, and the qubit state can be read out in a non-destructive manner with measurement times in the nanosecond regime, thereby allowing for the possibility of repeated measurements \cite{LupascuNatPhys2007}. Indeed, the observation of QZE was predicted for a single superconducting qubit using current technology \cite{MatsuzakiPRB2010}. Since dephasing times for superconducting qubits are around $1$ $\mu$s \cite{superconducting}, and both the dephasing time and the measurement rate are expected to increase in the years to come, the experimental observation of the crossover from QZE to QAZE appears promising.
For many two-level systems, we propose to use a two-component Bose-Einstein condensate which interacts with a thermal reservoir via collisions, with the system Hamiltonian being $\delta J_x$, where the energy bias can be set to be negligible \cite{KurizkiPRL2011}. If intermode collisions dominate, then we realize the same system-Hamiltonian as before, up to a unitary transformation. The state that we can repeatedly prepare now is the state with maximum population difference between the two modes. Measurement of the population difference corresponds to measurement of $J_z$, and it may be achieved by a non-destructive measurement technique, such as phase-contrast imaging \cite{HigbiePRL2005}.

{\it Conclusion} ---In conclusion, we have shown that even the simple single-spin pure dephasing model can manifest both the Zeno and anti-Zeno effects, with the transition between these two regimes strongly dependent on the environment properties. Multiple (local) transitions between Zeno and anti-Zeno regimes may occur for many spins coupled to a common environment. Finally, the disturbance to the environment by repeated measurements
 is shown to have a non-negligible influence on the Zeno and anti-Zeno effects.  Experimental studies of these effects, though challenging, should be an important step towards measurement-based quantum control in open quantum systems.

A.Z.C. is supported by the Singapore National Research Foundation under NRF Grant No.~NRF-NRFF2011-07.





%

\pagebreak

\onecolumngrid

\setcounter{equation}{0}

\section*{Supplementary Material: Zeno and anti-Zeno effects on Dephasing}

In this Supplementary material, we use the same notation as introduced in our main text.

\subsection*{Comparison between dissipative model and the dephasing model}

We start by briefly recapping the theoretical derivation of the quantum Zeno and anti-Zeno effects in a system undergoing population decay. Consider a two-level system which is prepared in its excited state. Due to the coupling with the environment, the two-level system spontaneously decays to the ground state. This decay process can be modelled using the Hamiltonian,
\begin{equation}
\label{decayHamiltonian}
H = \frac{\omega_0}{2}\sigma_z + \sum_k \omega_k b_k^\dagger b_k + \sum_k (g_k^* b_k \sigma^{+} + g_k b_k^\dagger \sigma^{-}),
\end{equation}
where $\omega_0$ is the energy difference between the two levels, $\sigma_z$ is the standard Pauli matrix while $\sigma_{+}$ and $\sigma_{-}$ are the raising and lowering operators, and $b_k$ and $b_k^\dagger$ are the annihilation and creation operators for mode $k$ of the environment. Note that the rotating-wave approximation has been made. Starting from the state $\ket{e,0}$, which means that the atom is in the excited state and that the environment is in the vacuum state,
the system-environment state at time $t$ can be written as
\begin{align}
\ket{\psi(t)} &= e^{-iHt} \ket{e,0} \notag \\
=& f(t) \ket{e,0} + \sum_k f_k(t) \ket{g,k},
\end{align}
where $\ket{g,k}$ means that the atom is in the ground state and that mode $k$ of the environment is excited. It is a simple exercise in first-order time-dependent perturbation theory to show that
\begin{equation}
f_k(t) = -ig_k e^{-i(\omega_k + \omega_0)t/2} t \, \text{sinc} [(\omega_k - \omega_0)t/2].
\end{equation}
We can then find the survival probability, defined as the probability that the atom is still in its excited state (using first-order time-dependent perturbation theory) as the following:
\begin{equation*}
s(t) = 1 - t^2 \sum_k |g_k|^2 \text{sinc}^2 [(\omega_k - \omega_0)t/2].
\end{equation*}
After $N$ measurements, each performed after a time interval $\tau$, the survival probability is
\begin{equation}
S(t = N\tau) = [s(\tau)]^N \equiv e^{-\Gamma(\tau)t_N},
\end{equation}
for total time $t_N = N\tau$. We then obtain the decay rate as (see Ref.~\cite{KurizkiNature2000} of main text)
\begin{align}
\label{originaldecayrate}
\Gamma(\tau) = -\frac{1}{\tau} \ln s(\tau) = \tau \sum_k |g_k|^2 \text{sinc}^2 [(\omega_k - \omega_0)\tau/2].
\end{align}
This decay rate, modified due to the repeated measurements, is thus dependent on the overlap between the environment spectral density and a measurement-dependent widening of the two-level system energy difference.

\begin{figure}[t]
\centering
\includegraphics[scale = 0.7]{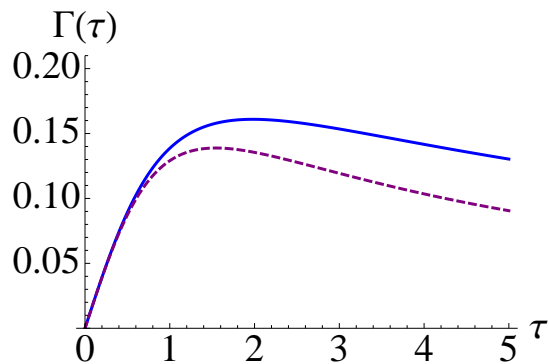}
\caption{(color online) Behaviour of $\Gamma(\tau)$ as a function of $\tau$ using Eq.~\eqref{originaldecayrate} (solid, blue) and using Eq.~\eqref{Gammadephasing} (dashed, purple) for $\omega_0 = 0$, $\beta \rightarrow \infty$, $\theta = \pi/2$ and $\phi = 0$. Here we have used $G = 0.2$ and $\omega_c = 1$.}
\label{nonRWA}
\end{figure}

\begin{figure}[t]
\centering
\includegraphics[scale = 0.7]{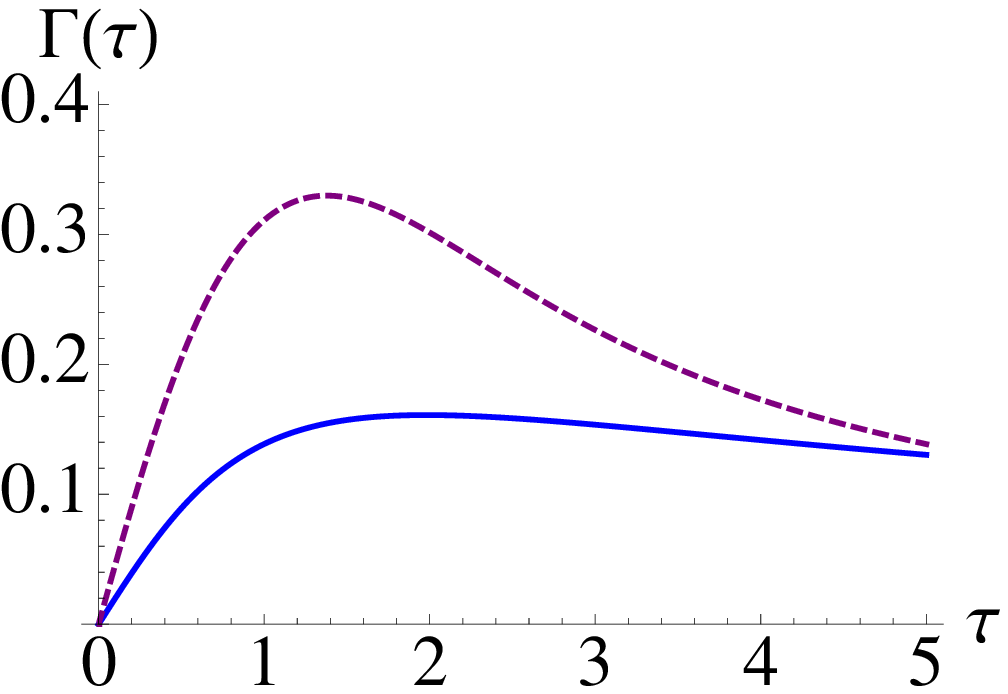}
\caption{(color online) Behaviour of $\Gamma(\tau)$ as a function of $\tau$ using Eq.~\eqref{originaldecayrate} (solid, blue) and using Eq.~\eqref{Gammadephasing} (dashed, purple) for $\beta = 1$. The rest of the parameters used are the same as Fig.~\ref{nonRWA}.}
\label{labelfinitetemperature}
\end{figure}

\begin{figure}[t]
\centering
\includegraphics[scale = 0.7]{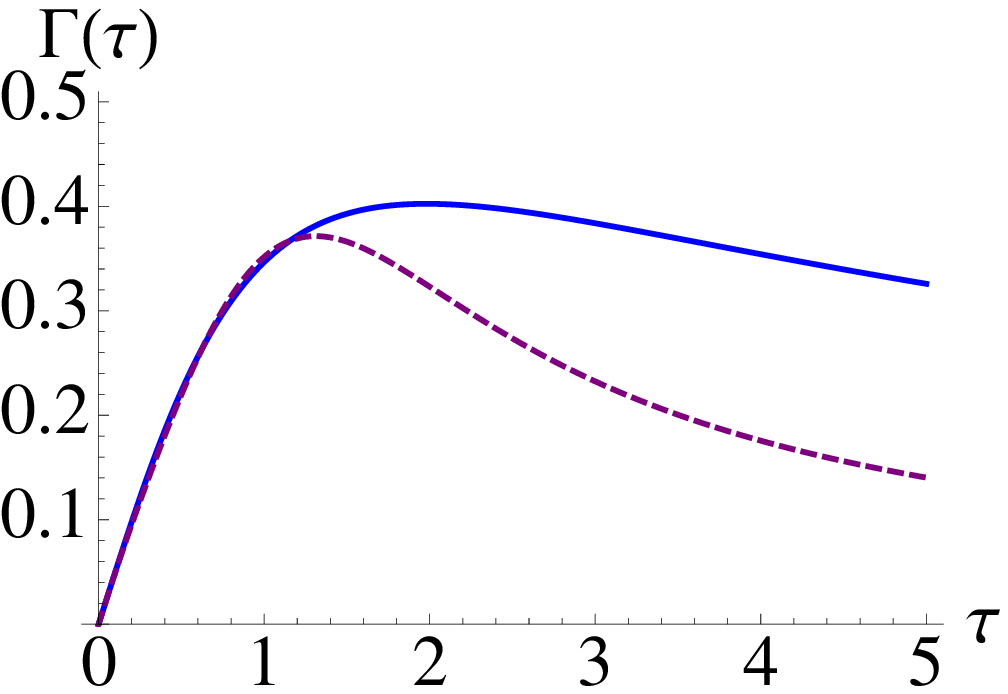}
\caption{(color online) Behaviour of $\Gamma(\tau)$ as a function of $\tau$ using Eq.~\eqref{originaldecayrate} (solid, blue) and using $\Gamma(\tau,N)$ with $N = 3$ (dashed, purple) for $G = 0.5$. The rest of the parameters used are the same as Fig.~\ref{nonRWA}.}
\label{strongercoupling}
\end{figure}

\begin{figure}[t]
\centering
\includegraphics[scale = 0.7]{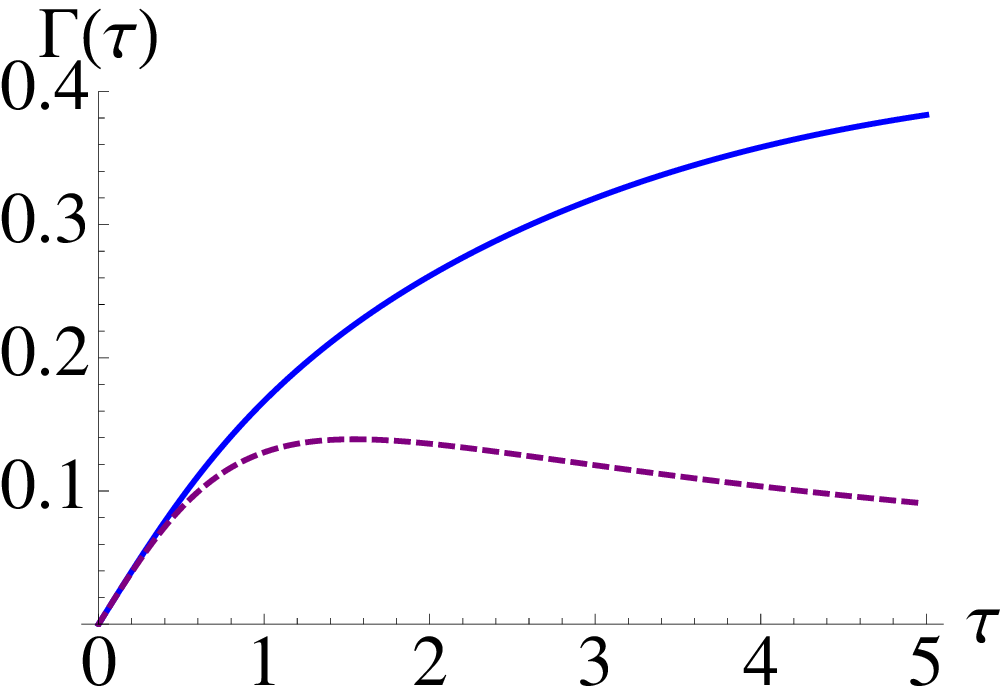}
\caption{(color online) Behaviour of $\Gamma(\tau)$ as a function of $\tau$ using Eq.~\eqref{originaldecayrate} (solid, blue) and using Eq.~\eqref{Gammadephasing} (dashed, purple) for $\omega_0 = 1$. The rest of the parameters used are the same as Fig.~\ref{nonRWA}.}
\label{finitebias}
\end{figure}

\begin{figure}[t]
\centering
\includegraphics[scale = 0.7]{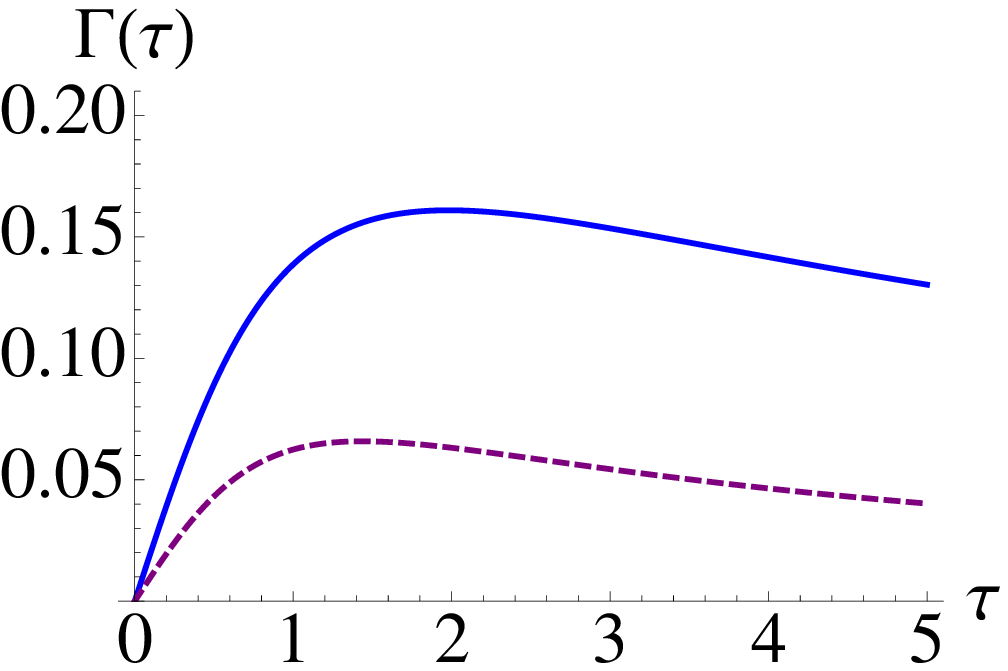}
\caption{(color online) Behaviour of $\Gamma(\tau)$ as a function of $\tau$ using Eq.~\eqref{originaldecayrate} (solid, blue) and using Eq.~\eqref{Gammadephasing} (dashed, purple) for $\theta = \pi/4$. The rest of the parameters used are the same as Fig.~\ref{nonRWA}.}
\label{thetapiby4}
\end{figure}

Let us now look at the pure dephasing model given by
\begin{equation}
\label{dephasingHamiltonian}
H = \frac{\omega_0}{2}\sigma_z + \sum_k \omega_k b_k^\dagger b_k + \sigma_z \sum_k (g_k^* b_k + g_k b_k^\dagger).
\end{equation}
As explained in the main text, if we assume that the state of the environment changes negligibly between each measurement, we have
\begin{equation}
\label{Gammadephasing}
\Gamma(\tau) = -\frac{1}{\tau} \ln \lbrace 1 - \frac{1}{2} \sin^2 \theta [1 - e^{-\gamma(\tau)}]\rbrace,
\end{equation}
where we repeatedly prepare the state $\ket{\psi} = \cos\left(\frac{\theta}{2}\right) \ket{e} + e^{i\phi}\sin\left(\frac{\theta}{2}\right) \ket{g}$, and we have removed the evolution due to the system's own Hamiltonian $H_S=\frac{\omega_0}{2}\sigma_z$ before each measurement via a unitary rotation. Clearly, Eq.~\eqref{originaldecayrate} and Eq.~\eqref{Gammadephasing} are different.

Now if $\omega_0 = 0$, $\beta \rightarrow \infty$, $\theta = \pi/2$ and $\phi = 0$, we can rotate the Hamiltonian Eq.~\eqref{dephasingHamiltonian} about the $y$-axis and obtain
$$H =\sum_k \omega_k b_k^\dagger b_k + \sigma_x \sum_k (g_k^* b_k + g_k b_k^\dagger).$$
Even this does not describe the same physics as Eq.~\eqref{decayHamiltonian} with $\omega_0 = 0$ since rotating-wave approximation has been made in obtaining Eq.~\eqref{decayHamiltonian}. The effect of these extra terms is illustrated in Fig.~\ref{nonRWA}. Furthermore, for finite temperature, Eq.~\eqref{Gammadephasing} becomes more different from Eq.~\eqref{originaldecayrate} because the latter is for zero temperature [see Fig.~\ref{labelfinitetemperature}]. At the same zero temperature, Eqs.~\eqref{Gammadephasing} and \eqref{originaldecayrate} are also much different for moderate coupling strength [see Fig.~\ref{strongercoupling}], where we have taken system-environment correlations into account.

What happens for finite $\omega_0$? In this case, it is important to realize that for the decay model, whether or not we apply $U_R(\tau) = e^{iH_S\tau}$ makes no difference on the decay rate and thus Eq.~\eqref{originaldecayrate} still depends on $\omega_0$. This is because we project onto an eigenstate of $H_S$, so if $U_R(\tau)$ is present, it only leads to a phase factor, which gets removed when we calculate the modulus squared to find probabilities. On the other hand, applying $U_R(\tau)$ makes Eq.~\eqref{Gammadephasing} independent of $\omega_0$. Thus, if $\omega_0 \neq 0$, we again have a difference, as shown in Fig.~\ref{finitebias}.

Finally, for a different state preparation, the two expressions lead to differences [see Fig.~\ref{thetapiby4}].

\subsection*{Effect of interaction on survival probability}

In this section, we look at what happens to the survival probability if our system Hamiltonian is
\begin{equation}
H = \chi J_z^2,
\end{equation}
and if there is no interaction with the environment. Such a Hamiltonian describes interaction between the two-level systems (see Ref.~\cite{GrossNature2010} in main text), in the same fashion as how a common environment induces indirect interaction between the two-level systems (for the environment-induced case, the indirect interaction is time-dependent). Here, purely to gain some qualitative understanding, we take $\chi$ to be time-independent. This is clearly a simplified case, but it suffices to illustrate the possible effect of an interaction induced by a common environment. Defining $s(\tau)$ to be the survival probability for a single measurement ($S(t = N\tau)$ can be written as approximately $[s(\tau)]^N$), it is simple to write for the spin coherent state defined in the main text that
\begin{equation}
s(\tau) = \left[ \frac{|\varsigma|}{(1 + |\varsigma|^2)}\right]^{4J} \sum_{m,n} |\varsigma|^{2(m + n)} \binom{2J}{J + m} \binom{2J}{J + n} e^{-i \chi \tau (m^2 - n^2)}.
\end{equation}

\begin{figure}[t]
\centering
\includegraphics[scale = 0.6]{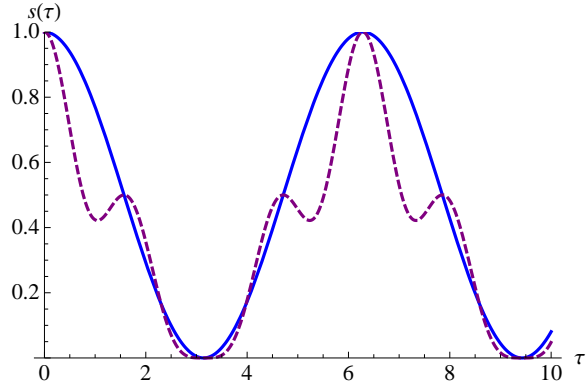}
\caption{(color online) Behaviour of the survival probability $s(\tau)$ as the interval $\tau$ changes for $J = 1$ (solid, blue) and $J = 2$ (dashed, purple). The interaction causes $s(\tau)$ to increase and decrease repeatedly. Here we have used $\theta = \pi/2$ and $\phi = 0$.}
\label{figsurvivalprobabilityinteraction}
\end{figure}

As shown in Fig.~\ref{figsurvivalprobabilityinteraction}, the interaction clearly causes the survival probability to increase and decrease repeatedly. It is precisely this effect that causes multiple Zeno and anti-Zeno regimes to emerge when the interaction is itself induced by the environment.

\subsection*{Changing state of environment between measurements}

We start by noting that the total density matrix just before the $N\textsuperscript{th}$ measurement is given by
\begin{align*}
\rho_{\text{tot}}(t \rightarrow N\tau) = \lbrace U(\tau)[P_{\psi}U(\tau)]^{N-1} \rho_{\text{tot}}(0) [U^\dagger(\tau)P_{\psi}]^{N-1}U^\dagger(\tau)  \rbrace /Z_{N-1},
\end{align*}
with $Z_{N-1}$ the normalization factor. Explicitly,
\begin{equation*}
Z_{N - 1} = \text{Tr}_{S,B} \lbrace [P_{\psi} U(\tau)]^{N-1} \rho_{\text{tot}}(0) [U^\dagger(\tau)P_{\psi}]^{N-1} \rbrace,
\end{equation*}
where $\text{Tr}_{S,B}$ denotes taking trace over $S$ (the system) and $B$ (the environment) and cyclic invariance of the trace has been used. Note that $Z_0 = 1$ because we take $\rho_{\text{tot}}(0)$ to be normalized. Now $\rho_{\text{tot}}(t\rightarrow N\tau)$ can be used to calculate the `success' probability (where `success' means that we measure the system state to be $\ket{\psi}$) for the $N\textsuperscript{th}$ measurement provided that all the previous measurements have been a success as well. This probability is 
\begin{equation}
\text{Tr}_{S,B} \lbrace P_{\psi} U(\tau) [P_{\psi} U(\tau)]^{N-1} \rho_{\text{tot}}(0) [U^\dagger(\tau)P_{\psi}]^{N-1} U^\dagger(\tau) \rbrace /Z_{N-1}. \nonumber
\end{equation}
Using the fact that $P_{\psi}^2 = P_{\psi}$, and the cyclic invariance again, we can write this the probability of success of the $N$th measurement with all previous measurements successful as
\begin{equation}
\text{Tr}_{S,B} \lbrace [P_{\psi} U(\tau)]^{N} \rho_{\text{tot}}(0) [U^\dagger(\tau)P_{\psi}]^{N} \rbrace /Z_{N-1}. \nonumber
\end{equation}
Thus, this makes it clear that the success probability of the $N$th measurement with all previous measurements successful is simply $\frac{Z_N}{Z_{N-1}}$.
Now we can find $S(t = N\tau)$ which is the probability that all the measurements are successful. This is then equal to
\begin{equation*}
S(t = N\tau) = \frac{Z_N}{Z_{N-1}} \frac{Z_{N-1}}{Z_{N-2}} \hdots \frac{Z_1}{Z_0} = \frac{Z_N}{Z_0} = Z_N.
\end{equation*}
So we finally have that
\begin{equation*}
S(t = N\tau) = \text{Tr}_{S,B} \lbrace [P_{\psi} U(\tau)]^{N} \rho_{\text{tot}}(0) [U^\dagger(\tau)P_{\psi}]^{N}\rbrace.
\end{equation*}
Next, we assume that $\rho_{\text{tot}}(0) = P_{\psi} \otimes \rho_B$. Since we are already focusing on how (multiple) measurements can disturb the state of the environment, this initial-product-state assumption is just a convenient starting point for theoretical considerations (that is, at time zero we prepare the system on state $|\psi\rangle$ and we assume that the environment is at its equilibrium state).  Then we find that
\begin{equation*}
S(t = N\tau) = \text{Tr}_{B} \lbrace \rho_B [\opav{\psi}{U^\dagger(\tau)}{\psi}]^{N} [\opav{\psi}{U(\tau)}{\psi}]^{N}\rbrace.
\end{equation*}
To proceed, we note that
\begin{equation*}
\opav{\psi}{U(\tau)}{\psi} = e^{-iH_B \tau} X(\tau),
\end{equation*}
where
\begin{equation*}
X(\tau) = \sum_l |\ip{l}{\psi}|^2 e^{-il^2\Delta(\tau)} e^{lR(\tau)},
\end{equation*}
$R(\tau) = \sum_k [\alpha_k(\tau) b_k^\dagger - \alpha_k^* (\tau) b_k]$, $\alpha_k(\tau) = \frac{2g_k}{\omega_k}(1 - e^{i\omega_k \tau})$, and $l$ labels the $J_z$ eigenstates. Now we can write
\begin{equation*}
S(t = N\tau) = \text{Tr}_B \lbrace \rho_B [X^\dagger (\tau) e^{iH_B \tau}]^N [e^{-iH_B \tau} X(\tau)]^N \rbrace.
\end{equation*}
This is still unwieldy, but using the fact that $H_B$ commutes with $\rho_B$, it is easy to show that
\begin{equation*}
S(t = N\tau) = \text{Tr}_B \lbrace \rho_B X_1^\dagger (\tau) X_2^\dagger (\tau) \hdots X_N^\dagger (\tau) X_N(\tau) X_{N-1}(\tau) \hdots X_1(\tau)\rbrace,
\end{equation*}
where
\begin{equation*}
X_p(\tau) = e^{ipH_B \tau} X(\tau) e^{-ipH_B \tau} =  \sum_l e^{-i \Delta(\tau) l^2} e^{lR_p(\tau)} |\ip{l}{\psi}|^2,
\end{equation*}
with $R_p(\tau) = \sum_k [\alpha_{k,p}(\tau) b_k^\dagger - \alpha_{k,p}^* (\tau) b_k]$ where $\alpha_{k,p}(\tau) = \frac{2g_k}{\omega_k} e^{ip\omega_k \tau} (1 - e^{i\omega_k \tau})$.
To proceed further, we need to combine $X_1^\dagger (\tau) \hdots X_N^\dagger (\tau) X_N(\tau) \hdots X_1(\tau)$ into one operator, following which we can use the identity $\text{Tr}_B \lbrace \rho_B e^A \rbrace = e^{\langle A^2 \rangle_B/2}$, where $\langle A^2 \rangle_B = \text{Tr}_B \lbrace \rho_B A^2 \rbrace$ and $A$ is a linear combination of annihilation and creation operators. However, $R_p(\tau)$ does not commute with $R_{p'}(\tau)$ for $p \neq p'$. Fortunately, we find that
$[R_p(\tau), R_{p'}(\tau)] = 4i\mu_{pp'}(\tau)$, where
\begin{equation*}
\mu_{pp'}(\tau) = \sum_k \frac{4|g_k|^2}{\omega_k^2} [1 - \cos(\omega_k \tau)] \sin[(p - p')\omega_k \tau].
\end{equation*}
Since this is a c-number, $e^{l_p R_p(\tau)}$ multiplied by $e^{l_{p'} R_{p'}(\tau)}$ can be written as a single exponential along with a phase factor. Once these phase factors are taken into account, what is left to calculate is of the form
\begin{equation*}
\text{Tr}_B \lbrace \rho_B e^{(l_1 - l_1')R_1(\tau) + (l_2 - l_2')R_2(\tau) + \hdots + (l_N - l_N')R_N(\tau)}\rbrace.
\end{equation*}
This can be done by using the identities
\begin{align*}
&\langle R_m^2(\tau) \rangle_B = 2\gamma(\tau) \notag \\
&\langle R_m(\tau) R_n(\tau) + R_n(\tau) R_m(\tau) \rangle_B = 4\gamma_{mn}(\tau) =  \notag \\
&4\sum_{k} \frac{4|g_k|^2}{\omega_k^2} [1 - \cos(\omega_k \tau)] \cos[(m - n)\omega_k \tau] \coth\left( \frac{\beta \omega_k}{2} \right),
\end{align*}
We can now put these pieces together and write down the survival probability for different $N$. For example, for $N = 2$, we have
\begin{align}
&S(t = 2\tau) = \sum_{l_1 l_2} \sum_{l_1' l_2'} e^{-i\Delta(\tau)(l_1^2 + l_2^2 - l_1'^2 - l_2'^2)}
|\ip{l_1}{\psi}|^2 |\ip{l_2}{\psi}|^2 |\ip{l_1'}{\psi}|^2 |\ip{l_2'}{\psi}|^2 \times \notag \\
&e^{-(l_1 - l_1')^2 \gamma(\tau)} e^{-(l_2 - l_2')^2 \gamma(\tau)}
e^{-2(l_1 - l_1')(l_2 - l_2')\gamma_{21}(\tau)} e^{i2\mu_{21}(\tau)(l_1 l_2 + l_1'l_2 - l_1 l_2' - l_1' l_2')}.
\end{align}
In an analogous manner, expressions can be written for larger $N$. For $N = 3$, we have
\begin{align}
&S(t = 3\tau) = \sum_{l_1 l_2 l_3} \sum_{l_1' l_2' l_3'} e^{-i\Delta(\tau)(l_1^2 + l_2^2 + l_3^2 - l_1'^2 - l_2'^2 - l_3'^2)}
|\ip{l_1}{\psi}|^2 |\ip{l_2}{\psi}|^2 |\ip{l_3}{\psi}|^2 |\ip{l_1'}{\psi}|^2 |\ip{l_2'}{\psi}|^2 |\ip{l_3'}{\psi}|^2\times \notag \\
&e^{-(l_1 - l_1')^2 \gamma(\tau)} e^{-(l_2 - l_2')^2 \gamma(\tau)} e^{-(l_3 - l_3')^2 \gamma(\tau)}
e^{-2(l_1 - l_1')(l_2 - l_2')\gamma_{21}(\tau)} e^{-2(l_1 - l_1')(l_3 - l_3')\gamma_{31}(\tau)} e^{-2(l_2 - l_2')(l_3 - l_3')\gamma_{32}(\tau)} \times \notag \\
&e^{i2\mu_{21}(\tau)(l_1 l_2 + l_1'l_2 - l_1 l_2' - l_1' l_2')} e^{i2\mu_{31}(\tau)(l_1 l_3 + l_1'l_3 - l_1 l_3' - l_1' l_3')} e^{i2\mu_{32}(\tau)(l_2 l_3 + l_2'l_3 - l_2 l_3' - l_2' l_3')}.
\end{align}
One can generalize the expression for any $N$. Let us take $N = 5$. Then we have ten indices to sum over: $l_1 \hdots l_5$ and $l_1' \hdots l_5'$. So write down the summation signs with these indices. Next comes the factor due to the indirect interactions. This is easy: it is simply $e^{-i\Delta(\tau)(l_1^2 + \hdots + l_5^2 - l_1'^2 - \hdots - l_5'^2)}$. Next come the factors due to the projections. Again, this is easy: we will get $|\ip{l_1}{\psi}|^2 \hdots |\ip{l_5}{\psi}|^2 |\ip{l_1'}{\psi}|^2 \hdots |\ip{l_5'}{\psi}|^2$. Then, put in the factors $e^{-(l_j - l_j')^2 \gamma(\tau)}$ for $j$ from $1$ to $5$. Next, take the numbers $1$ to $5$ and form pairs from them of the form $(j,k)$, with $j > k$. Use these pairs to write down the factors $e^{-2(l_k - l_k')(l_j - l_j')\gamma_{jk}(\tau)}$ for all the pairs. Finally, put in the factors $e^{i2\mu_{jk}(\tau)(l_k l_j + l_k' l_j - l_k l_j' - l_k' l_j')}$ for all the pairs. Following this recipe, we can write down the survival probability for any $N$. These `pair' factors essentially take into account the fact that the decoherence of the system due to the environment   changes as a result of the changing state of the environment caused by the correlations and measurements.


\begin{thebibliography}{99}%
\makeatletter
\providecommand \@ifxundefined [1]{%
 \@ifx{#1\undefined}
}%
\providecommand \@ifnum [1]{%
 \ifnum #1\expandafter \@firstoftwo
 \else \expandafter \@secondoftwo
 \fi
}%
\providecommand \@ifx [1]{%
 \ifx #1\expandafter \@firstoftwo
 \else \expandafter \@secondoftwo
 \fi
}%
\providecommand \natexlab [1]{#1}%
\providecommand \enquote  [1]{``#1''}%
\providecommand \bibnamefont  [1]{#1}%
\providecommand \bibfnamefont [1]{#1}%
\providecommand \citenamefont [1]{#1}%
\providecommand \href@noop [0]{\@secondoftwo}%
\providecommand \href [0]{\begingroup \@sanitize@url \@href}%
\providecommand \@href[1]{\@@startlink{#1}\@@href}%
\providecommand \@@href[1]{\endgroup#1\@@endlink}%
\providecommand \@sanitize@url [0]{\catcode `\\12\catcode `\$12\catcode
  `\&12\catcode `\#12\catcode `\^12\catcode `\_12\catcode `\%12\relax}%
\providecommand \@@startlink[1]{}%
\providecommand \@@endlink[0]{}%
\providecommand \url  [0]{\begingroup\@sanitize@url \@url }%
\providecommand \@url [1]{\endgroup\@href {#1}{\urlprefix }}%
\providecommand \urlprefix  [0]{URL }%
\providecommand \Eprint [0]{\href }%
\providecommand \doibase [0]{http://dx.doi.org/}%
\providecommand \selectlanguage [0]{\@gobble}%
\providecommand \bibinfo  [0]{\@secondoftwo}%
\providecommand \bibfield  [0]{\@secondoftwo}%
\providecommand \translation [1]{[#1]}%
\providecommand \BibitemOpen [0]{}%
\providecommand \bibitemStop [0]{}%
\providecommand \bibitemNoStop [0]{.\EOS\space}%
\providecommand \EOS [0]{\spacefactor3000\relax}%
\providecommand \BibitemShut  [1]{\csname bibitem#1\endcsname}%
\let\auto@bib@innerbib\@empty
\bibitem [{\citenamefont {Misra}\ and\ \citenamefont
  {Sudarshan}(1977)}]{Sudarshan1977}%
  \BibitemOpen
  \bibfield  {author} {\bibinfo {author} {\bibfnamefont {B.}~\bibnamefont
  {Misra}}\ and\ \bibinfo {author} {\bibfnamefont {E.~C.~G.}\ \bibnamefont
  {Sudarshan}},\ }\href@noop {} {\bibfield  {journal} {\bibinfo  {journal} {J.
  Math. Phys. (N. Y.)}\ }\textbf {\bibinfo {volume} {18}},\ \bibinfo {pages}
  {756} (\bibinfo {year} {1977})}\BibitemShut {NoStop}%
  \bibitem{QZDreferences} P.~Facchi, V.~Gorini, G.~Marmo, S.~Pascazio and E.~C.~G.~Sudarshan, Phys.~Lett.~A~{\textbf{275}}, 12 (2000); P.~Facchi and S.~Pascazio, Phys.~Rev.~Lett.~{\textbf{89}}, 080401 (2002); P.~Facchi and S.~Pascazio, J.~Phys.~A: Math.~Theor.~{\textbf{41}}, 493001 (2008); X.~B.~Wang, J.~Q.~You and F.~Nori, Phys.~Rev.~A~{\textbf{77}}, 062339 (2008); S.~Maniscalco, F.~Francica, R.~L.~Zaffino, N.~L.~Gullo and F.~Plastina, Phys.~Rev.~Lett.~{\textbf{100}}, 090503 (2008); P.~Facchi and M.~Ligab\`{o}, J.~Math.~Phys.~{\textbf{51}}, 022103 (2010); J.~M.~Raimond, P.~Facchi, B.~Peaudecerf, S.~Pascazio, C.~Sayrin, I.~Dotsenko, S.~Gleyzes, M.~Brune, and S.~Haroche, Phys.~Rev.~A~{\textbf{86}}, 032120 (2012); A.~ Signoles, A.~Facon, D.~Grosso, I.~Dotsenko, S.~Haroche, J.~M.~Raimond, M.~Brune and S.~Gleyzes, e-print arXiv:1402.0111 (2014).

  \bibitem [{\citenamefont {Kofman}\ and\ \citenamefont
  {Kurizki}(2000)}]{KurizkiNature2000}%
  \BibitemOpen
  \bibfield  {author} {\bibinfo {author} {\bibfnamefont {A.~G.}\ \bibnamefont
  {Kofman}}\ and\ \bibinfo {author} {\bibfnamefont {G.}~\bibnamefont
  {Kurizki}},\ }\href {\doibase 10.1038/35014537} {\bibfield  {journal}
  {\bibinfo  {journal} {Nature (London)}\ }\textbf {\bibinfo {volume} {405}},\
  \bibinfo {pages} {546} (\bibinfo {year} {2000})}\BibitemShut {NoStop}%
\bibitem [{\citenamefont {Fischer}\ \emph {et~al.}(2001)\citenamefont
  {Fischer}, \citenamefont {Guti\'errez-Medina},\ and\ \citenamefont
  {Raizen}}]{RaizenPRL2001}%
  \BibitemOpen
  \bibfield  {author} {\bibinfo {author} {\bibfnamefont {M.~C.}\ \bibnamefont
  {Fischer}}, \bibinfo {author} {\bibfnamefont {B.}~\bibnamefont
  {Guti\'errez-Medina}}, \ and\ \bibinfo {author} {\bibfnamefont {M.~G.}\
  \bibnamefont {Raizen}},\ }\href {\doibase 10.1103/PhysRevLett.87.040402}
  {\bibfield  {journal} {\bibinfo  {journal} {Phys. Rev. Lett.}\ }\textbf
  {\bibinfo {volume} {87}},\ \bibinfo {pages} {040402} (\bibinfo {year}
  {2001})}\BibitemShut {NoStop}%
\bibitem [{\citenamefont {Barone}\ \emph {et~al.}(2004)\citenamefont {Barone},
  \citenamefont {Kurizki},\ and\ \citenamefont {Kofman}}]{BaronePRL2004}%
  \BibitemOpen
  \bibfield  {author} {\bibinfo {author} {\bibfnamefont {A.}~\bibnamefont
  {Barone}}, \bibinfo {author} {\bibfnamefont {G.}~\bibnamefont {Kurizki}}, \
  and\ \bibinfo {author} {\bibfnamefont {A.~G.}\ \bibnamefont {Kofman}},\
  }\href {\doibase 10.1103/PhysRevLett.92.200403} {\bibfield  {journal}
  {\bibinfo  {journal} {Phys. Rev. Lett.}\ }\textbf {\bibinfo {volume} {92}},\
  \bibinfo {pages} {200403} (\bibinfo {year} {2004})}\BibitemShut {NoStop}%
\bibitem [{\citenamefont {Fujii}\ and\ \citenamefont
  {Yamamoto}(2010)}]{YamamotoPRA2010}%
  \BibitemOpen
  \bibfield  {author} {\bibinfo {author} {\bibfnamefont {K.}~\bibnamefont
  {Fujii}}\ and\ \bibinfo {author} {\bibfnamefont {K.}~\bibnamefont
  {Yamamoto}},\ }\href {\doibase 10.1103/PhysRevA.82.042109} {\bibfield
  {journal} {\bibinfo  {journal} {Phys. Rev. A}\ }\textbf {\bibinfo {volume}
  {82}},\ \bibinfo {pages} {042109} (\bibinfo {year} {2010})}\BibitemShut
  {NoStop}%
\bibitem [{\citenamefont {Chen}\ \emph {et~al.}(2010)\citenamefont {Chen},
  \citenamefont {Tsai},\ and\ \citenamefont {Bennett}}]{BennettPRB2010}%
  \BibitemOpen
  \bibfield  {author} {\bibinfo {author} {\bibfnamefont {P.-W.}\ \bibnamefont
  {Chen}}, \bibinfo {author} {\bibfnamefont {D.-B.}\ \bibnamefont {Tsai}}, \
  and\ \bibinfo {author} {\bibfnamefont {P.}~\bibnamefont {Bennett}},\ }\href
  {\doibase 10.1103/PhysRevB.81.115307} {\bibfield  {journal} {\bibinfo
  {journal} {Phys. Rev. B}\ }\textbf {\bibinfo {volume} {81}},\ \bibinfo
  {pages} {115307} (\bibinfo {year} {2010})}\BibitemShut {NoStop}%
\bibitem [{\citenamefont {Koshino}\ and\ \citenamefont
  {Shimizu}(2005)}]{KoshinoPhysRep2005}%
  \BibitemOpen
  \bibfield  {author} {\bibinfo {author} {\bibfnamefont {K.}~\bibnamefont
  {Koshino}}\ and\ \bibinfo {author} {\bibfnamefont {A.}~\bibnamefont
  {Shimizu}},\ }\href@noop {} {\bibfield  {journal} {\bibinfo  {journal} {Phys.
  Rep.}\ }\textbf {\bibinfo {volume} {412}},\ \bibinfo {pages} {191} (\bibinfo
  {year} {2005})}\BibitemShut {NoStop}%
\bibitem [{\citenamefont {Maniscalco}\ \emph {et~al.}(2006)\citenamefont
  {Maniscalco}, \citenamefont {Piilo},\ and\ \citenamefont
  {Suominen}}]{ManiscalcoPRL2006}%
  \BibitemOpen
  \bibfield  {author} {\bibinfo {author} {\bibfnamefont {S.}~\bibnamefont
  {Maniscalco}}, \bibinfo {author} {\bibfnamefont {J.}~\bibnamefont {Piilo}}, \
  and\ \bibinfo {author} {\bibfnamefont {K.-A.}\ \bibnamefont {Suominen}},\
  }\href {\doibase 10.1103/PhysRevLett.97.130402} {\bibfield  {journal}
  {\bibinfo  {journal} {Phys. Rev. Lett.}\ }\textbf {\bibinfo {volume} {97}},\
  \bibinfo {pages} {130402} (\bibinfo {year} {2006})}\BibitemShut {NoStop}%
\bibitem [{\citenamefont {Segal}\ and\ \citenamefont
  {Reichman}(2007)}]{SegalPRA2007}%
  \BibitemOpen
  \bibfield  {author} {\bibinfo {author} {\bibfnamefont {D.}~\bibnamefont
  {Segal}}\ and\ \bibinfo {author} {\bibfnamefont {D.~R.}\ \bibnamefont
  {Reichman}},\ }\href {\doibase 10.1103/PhysRevA.76.012109} {\bibfield
  {journal} {\bibinfo  {journal} {Phys. Rev. A}\ }\textbf {\bibinfo {volume}
  {76}},\ \bibinfo {pages} {012109} (\bibinfo {year} {2007})}\BibitemShut
  {NoStop}%
\bibitem [{\citenamefont {Zheng}\ \emph {et~al.}(2008)\citenamefont {Zheng},
  \citenamefont {Zhu},\ and\ \citenamefont {Zubairy}}]{ZhengPRL2008}%
  \BibitemOpen
  \bibfield  {author} {\bibinfo {author} {\bibfnamefont {H.}~\bibnamefont
  {Zheng}}, \bibinfo {author} {\bibfnamefont {S.~Y.}\ \bibnamefont {Zhu}}, \
  and\ \bibinfo {author} {\bibfnamefont {M.~S.}\ \bibnamefont {Zubairy}},\
  }\href {\doibase 10.1103/PhysRevLett.101.200404} {\bibfield  {journal}
  {\bibinfo  {journal} {Phys. Rev. Lett.}\ }\textbf {\bibinfo {volume} {101}},\
  \bibinfo {pages} {200404} (\bibinfo {year} {2008})}\BibitemShut {NoStop}%
\bibitem{AiPRA2010} Q.~Ai, Y.~Li, H.~Zheng and C.~P.~Sun, \pra\jn{81}, 042116 (2010).
\bibitem{ThilagamJMP2010} A.~Thilagam, J.~Phys.~A: Math.~Theor.~{\textbf{43}}, 155301 (2010).
\bibitem [{\citenamefont {Thilagam}(2013)}]{ThilagamJCP2013}%
  \BibitemOpen
  \bibfield  {author} {\bibinfo {author} {\bibfnamefont {A.}~\bibnamefont
  {Thilagam}},\ }\href@noop {} {\bibfield  {journal} {\bibinfo  {journal} {J.
  Chem. Phys.}\ }\textbf {\bibinfo {volume} {138}},\ \bibinfo {pages} {175102}
  (\bibinfo {year} {2013})}\BibitemShut {NoStop}%
\bibitem{NoteonZenotime} For example, it was calculated that for the 2P-1S transition in the hydrogen atom, measurements need to be performed is on the fs timescale [see P.~Facchi and S.~Pascazio, Phys.~Lett.~A~{\bf 241}, 139 (1998)].
\bibitem [{\citenamefont {Breuer}\ and\ \citenamefont
  {Petruccione}(2007)}]{BPbook}%
  \BibitemOpen
  \bibfield  {author} {\bibinfo {author} {\bibfnamefont {H.-P.}\ \bibnamefont
  {Breuer}}\ and\ \bibinfo {author} {\bibfnamefont {F.}~\bibnamefont
  {Petruccione}},\ }\href@noop {} {\emph {\bibinfo {title} {The Theory of Open
  Quantum Systems}}}\ (\bibinfo  {publisher} {Oxford University Press},\
  \bibinfo {address} {Oxford},\ \bibinfo {year} {2007})\BibitemShut {NoStop}%
\bibitem [{\citenamefont {Weiss}(2008)}]{Weissbook}%
  \BibitemOpen
  \bibfield  {author} {\bibinfo {author} {\bibfnamefont {U.}~\bibnamefont
  {Weiss}},\ }\href@noop {} {\emph {\bibinfo {title} {Quantum dissipative
  systems}}}\ (\bibinfo  {publisher} {World Scientific},\ \bibinfo {address}
  {Singapore},\ \bibinfo {year} {2008})\BibitemShut {NoStop}%
\bibitem{KurizkiNJP2010} See, for instance, G.~Gordon, D.~D.~B.~Rao and G.~Kurizki, New J.~Phys.~{\textbf{12}}, 053033 (2010).
\bibitem [{Sup()}]{Supple}%
  \BibitemOpen
  \href@noop {} {\bibinfo  {journal} {For more details on the comparison between the dissipative model and the dephasing model as well as the effect of indirect interactions and the system-environment correlations on the survival probability, refer to the
  Supplementary Material}\ }\BibitemShut {NoStop}%
\bibitem{Noteonlocal} `Local' QZE-QAZE transitions were considered for population decay in Refs.~\cite{KurizkiNature2000, SegalPRA2007, ThilagamJMP2010}. An alternative approach is to compare  the measurement modified decay rate with the decay rate without measurement [P.~Facchi, H.~Nakazato and S.~Pascazio, Phys.~Rev.~Lett.~{\textbf{86}}, 2699 (2001)].  Here we use the former approach.
\bibitem{NoriRepProg2011} I.~Buluta, S.~Ashhab and F.~Nori, Rep.~Prog.~Phys.~{\bf 74}, 104401 (2011).
\bibitem [{\citenamefont {Vorrath}\ and\ \citenamefont
  {Brandes}(2005)}]{VorrathPRL2005}%
  \BibitemOpen
\bibfield  {journal} {  }\bibfield  {author} {\bibinfo {author} {\bibfnamefont
  {T.}~\bibnamefont {Vorrath}}\ and\ \bibinfo {author} {\bibfnamefont
  {T.}~\bibnamefont {Brandes}},\ }\href {\doibase
  10.1103/PhysRevLett.95.070402} {\bibfield  {journal} {\bibinfo  {journal}
  {Phys. Rev. Lett.}\ }\textbf {\bibinfo {volume} {95}},\ \bibinfo {pages}
  {070402} (\bibinfo {year} {2005})}\BibitemShut {NoStop}%
\bibitem [{\citenamefont {Gross}\ \emph {et~al.}(2010)\citenamefont {Gross},
  \citenamefont {Zibold}, \citenamefont {Nicklas}, \citenamefont {Est{\`e}ve},\
  and\ \citenamefont {Oberthaler}}]{GrossNature2010}%
  \BibitemOpen
  \bibfield  {author} {\bibinfo {author} {\bibfnamefont {C.}~\bibnamefont
  {Gross}}, \bibinfo {author} {\bibfnamefont {T.}~\bibnamefont {Zibold}},
  \bibinfo {author} {\bibfnamefont {E.}~\bibnamefont {Nicklas}}, \bibinfo
  {author} {\bibfnamefont {J.}~\bibnamefont {Est{\`e}ve}}, \ and\ \bibinfo
  {author} {\bibfnamefont {M.~K.}\ \bibnamefont {Oberthaler}},\ }\href
  {\doibase 10.1038/nature08919} {\bibfield  {journal} {\bibinfo  {journal}
  {Nature}\ }\textbf {\bibinfo {volume} {464}},\ \bibinfo {pages} {1165}
  (\bibinfo {year} {2010})}\BibitemShut {NoStop}%
\bibitem [{\citenamefont {Riedel}\ \emph {et~al.}(2010)\citenamefont {Riedel},
  \citenamefont {B{\"o}hi}, \citenamefont {Li}, \citenamefont {H{\"a}nsch},
  \citenamefont {Sinatra},\ and\ \citenamefont {Treutlein}}]{RiedelNature2010}%
  \BibitemOpen
  \bibfield  {author} {\bibinfo {author} {\bibfnamefont {M.~F.}\ \bibnamefont
  {Riedel}}, \bibinfo {author} {\bibfnamefont {P.}~\bibnamefont {B{\"o}hi}},
  \bibinfo {author} {\bibfnamefont {Y.}~\bibnamefont {Li}}, \bibinfo {author}
  {\bibfnamefont {T.~W.}\ \bibnamefont {H{\"a}nsch}}, \bibinfo {author}
  {\bibfnamefont {A.}~\bibnamefont {Sinatra}}, \ and\ \bibinfo {author}
  {\bibfnamefont {P.}~\bibnamefont {Treutlein}},\ }\href {\doibase
  10.1038/nature08988} {\bibfield  {journal} {\bibinfo  {journal} {Nature}\
  }\textbf {\bibinfo {volume} {464}},\ \bibinfo {pages} {1170} (\bibinfo {year}
  {2010})}\BibitemShut {NoStop}%
\bibitem [{\citenamefont {Bar-Gill}\ \emph {et~al.}(2011)\citenamefont
  {Bar-Gill}, \citenamefont {Rao},\ and\ \citenamefont
  {Kurizki}}]{KurizkiPRL2011}%
  \BibitemOpen
  \bibfield  {author} {\bibinfo {author} {\bibfnamefont {N.}~\bibnamefont
  {Bar-Gill}}, \bibinfo {author} {\bibfnamefont {D.~D.~B.}\ \bibnamefont
  {Rao}}, \ and\ \bibinfo {author} {\bibfnamefont {G.}~\bibnamefont
  {Kurizki}},\ }\href {\doibase 10.1103/PhysRevLett.107.010404} {\bibfield
  {journal} {\bibinfo  {journal} {Phys. Rev. Lett.}\ }\textbf {\bibinfo
  {volume} {107}},\ \bibinfo {pages} {010404} (\bibinfo {year}
  {2011})}\BibitemShut {NoStop}%
\bibitem [{\citenamefont {Castin}\ and\ \citenamefont
  {Dalibard}(1997)}]{CastinPRA1997}%
  \BibitemOpen
  \bibfield  {author} {\bibinfo {author} {\bibfnamefont {Y.}~\bibnamefont
  {Castin}}\ and\ \bibinfo {author} {\bibfnamefont {J.}~\bibnamefont
  {Dalibard}},\ }\href {\doibase 10.1103/PhysRevA.55.4330} {\bibfield
  {journal} {\bibinfo  {journal} {Phys. Rev. A}\ }\textbf {\bibinfo {volume}
  {55}},\ \bibinfo {pages} {4330} (\bibinfo {year} {1997})}\BibitemShut
  {NoStop}%
\bibitem{ArecchiPRA1972} F.~T.~Arecchi, E.~Courtens, R.~Gilmore and H.~Thomas, \pra\jn{6}, 2211 (1972).
\bibitem [{\citenamefont {Kofman}\ and\ \citenamefont
  {Kurizki}(2004)}]{KurizkiPRL2004}%
  \BibitemOpen
  \bibfield  {author} {\bibinfo {author} {\bibfnamefont {A.~G.}\ \bibnamefont
  {Kofman}}\ and\ \bibinfo {author} {\bibfnamefont {G.}~\bibnamefont
  {Kurizki}},\ }\href {\doibase 10.1103/PhysRevLett.93.130406} {\bibfield
  {journal} {\bibinfo  {journal} {Phys. Rev. Lett.}\ }\textbf {\bibinfo
  {volume} {93}},\ \bibinfo {pages} {130406} (\bibinfo {year}
  {2004})}\BibitemShut {NoStop}%
  \bibitem{adamCJC} A.~Z.~Chaudhry and J.~B.~Gong, Can.~J.~Chem. {\bf 92}, 119 (2014);
  \pra~{\bf 87}, 012129 (2013); \pra~{\bf 88}, 052107 (2013).
\bibitem{superconducting} J.~Clarke and F.~K.~Wilhelm, Nature (London) {\bf 453}, 1031 (2008); J. Q. You and F. Nori, Nature (London) {\bf 474}, 589 (2011).
\bibitem{atom-cavity}See, for example, K. Hennessy {\it et al.},
  Nature (London) {\bf 445}, 896 (2007); G. G\"{u}nter {\it et al.}, Nature {\bf 458}, 178 (2009);
  A. Auer and G. Burkard, Phys.~Rev. B {\bf 85}, 235140 (2012).
\bibitem{GessnerNatPhys2014} See, for example, M.~Gessner, M.~Ramm, T.~Pruttivarasin, A.~Buchleitner, H-P.~Breuer and H.~H\"{a}ffner, Nat.~Phys.~{\bf 10}, 105 (2014) and references therein.
    \bibitem{taka1} A.~G.~Dijkstra and Y.~Tanimura, Phil.~Trans.~R.~Soc.~{\bf A 370}, 3658 (2012); J.~Phys.~Soc.~Japan {\bf 81}, 063301 (2012).
\bibitem{HurPRB2012} K.~Le Hur, Phys.~Rev.~B, {\bf 85}, 140506(R) (2012).
\bibitem{LupascuNatPhys2007} A.~Lupa\c{s}cu, S.~Saito, T.~Picot, P.~C.~de Groot, C.~J.~P.~M.~Harmans and J.~E.~Mooij, Nat.~Phys.~{\bf 3}, 119 (2007).
\bibitem{MatsuzakiPRB2010} Y.~Matsuzaki, S.~Saito, K~Kakuyanagi and K.~Semba, \prb\jn{82}, 180518(R) (2010).
\bibitem{HigbiePRL2005} See, for instance, J.~M.~Higbie, L.~E.~Sadler, S.~Inouye, A.~P.~Chikkatur, S.~R.~Leslie, K.~L.~Moore, V.~Savalli and D.~M.~Stamper-Kurn, \prl\jn{95}, 050401 (2005).

\end{thebibliography}
\end{document}